\documentclass[preprintnumbers,prd,onecolumn,floatfix,superscriptaddress, nofootinbib]{revtex4-2}
\usepackage{graphicx}
\usepackage{setspace}
\usepackage{subfig}
\usepackage{epsfig}
\usepackage{bm}
\usepackage{amssymb}
\usepackage{float}
\usepackage{amsmath}
\usepackage{dcolumn}
\usepackage[colorlinks]{hyperref}
\usepackage[usenames,dvipsnames]{color}
\usepackage{enumitem}

\hypersetup{ breaklinks=true, pdfstartview={FitH}, colorlinks=true, linkcolor=blue, citecolor=red, filecolor=magenta, urlcolor=blue, anchorcolor=green, linktocpage=true }

\begin{document}

\title{ Cosmic reverberations on a constrained $ f(Q,T) $-model of the Universe}
\author{Akanksha Singh}
\email{akanksha.ma19@nsut.ac.in}
\affiliation{Department of Mathematics, Netaji Subhas University of Technology, New Delhi-110 078, India}
\author{Shaily}
\email{shailytyagi.iitkgp@gmail.com}
\affiliation{Department of Mathematics, Netaji Subhas University of Technology, New Delhi-110 078, India}
\affiliation{School of Computer Science Engineering and Technology, Bennett University, Greater Noida, India}
\author{J.~K. Singh}
\email{jksingh@nsut.ac.in}
\affiliation{Department of Mathematics, Netaji Subhas University of Technology, New Delhi-110 078, India}
\author{Ertan Güdekli}
\email{	gudekli@istanbul.edu.tr}
\affiliation{Department of Physics, Istanbul University, 34452 Istanbul, Turkey}

\begin{abstract}

In this paper, we construct an isotropic cosmological model in the $ f(Q, T) $ theory of gravity in the frame of a flat FLRW spacetime being $ Q $ the non-metricity tensor and $ T $ the trace of the energy-momentum tensor. The gravity function is taken to be a quadratic equation, $ f(Q, T)=\zeta Q^2 + \gamma T $, where $ \zeta<0 $ and $ \gamma $ are the arbitrary constants. We constrain the model parameters $ \alpha $ and $ H_0 $ using the recent observational datasets: the Hubble dataset (OHD), the $ Pantheon $ dataset of $ 1048 $ points, and the joint dataset (OHD + $ Pantheon $). The universe model transitions from an early deceleration state to an acceleration in late times. This model also provides the ekpyrotic phase of the universe on the redshift $ z>12.32 $. In this model, the Big Bang is described as a collision of branes, and thus, the Big Bang is not the beginning of time. Before the Big Bang, there is an ekpyrotic phase with the equation of state $ \omega >> 1 $. In late times, the undeviating Hubble measurements reduce the $ H_0 $ tension in the reconstructed $ f(Q, T) $ function. Additionally, we study various physical parameters of the model. Finally, our model describes a quintessence dark energy model at later times.
 
\end{abstract}

\maketitle

\section{Introduction}\label{intro}

\qquad The human accomplishment to observe more clearly has led to the discovery of differences between theoretical models and empirical evidence these days. It is believed that the cosmos is in its expansion phase, and it is interesting to note that the universe had a fast expansion just after the Big Bang, referred to as the inflationary era. The literature has examined various directions to understand the phase-transition behavior from early-time deceleration to late-time acceleration. In the last three decades, cosmologists have made some surprising observations that have concluded that the expansion nature of the universe is behaving in an accelerating manner \cite{WMAP:2012nax, SDSS:2003eyi, SDSS:2004kqt, BOSS:2012tck, SDSS:2005xqv, Jain:2003tba, Planck:2018vyg}. Such observations completely changed the thinking of cosmologists, who then reviewed Einstein's theory, which contains the cosmological constant. This accelerating expansion of the universe indicates the presence of dark sectors. Until now, various cosmological models have been proposed that try to be a mock-up of this era, with others using different approaches like scalar fields, modified gravity in distinct forms, etc. 

\qquad When we deal with the theory of gravitation, there are many existing potential theories in which General Relativity (GR) is the foundation of these various conceivable non-linear gravity theories (NLG). Modified theories of gravity are nothing but some modifications, generalizations, and complications of Einstein's GR. However, these modified theories are not able to fully explain the mysterious questions. The theory of GR is indeed a milestone achievement for us. Some concepts of GR have given a new direction to describing the universe. Several observational phenomena like black holes, neutron stars, and gravitational waves were determined during the $ 20^{th} $ century. Although GR is highly tested, the door for alternatives to GR is always open. According to the probes, the behavior of $ 95\% $ of the universe's matter content is unknown. GR is not enough to describe gravity completely as it does not engage the local energy-momentum tensor (EMT) \cite{Carmeli1990, Einstein1992, Yilmaz1992}. This theory is not able to explain the reason for speeding up the expansion of the universe and is also inadequate to describe the most curious energy, i.e., dark energy, which covers around $ 70\% $ of the universe. Therefore, in the last few decades, modified theories of gravity have become more famous because of better explanations regarding dark energy, dark matter, the expanding nature of the universe, etc. The modified theory of gravity discusses the early universe and the late-time universe. The phase transition from early-time deceleration to late-time acceleration, along with variation from the non-phantom phase to the phantom phase, without involving exotic matter.

\qquad Many of the excellent works have been done in the field of modified theories of gravity, including the scale-independent theory of Weyl, Eddington's theory of connections \cite{Eddington1923}, the scalar-tensor theory \cite{Wagoner:1970vr, Singh:2010zzo, Yazadjiev:2016pcb, Quiros:2019ktw}, string theory \cite{Singh:2009zzb, Singh:2010zzp}, Brans-Dicke's theory \cite{Brans:1961sx, SolaPeracaula:2019zsl, Singh:2016eom, Ziaie:2024saq}, the higher-dimensional theory of Kaluza-Klein, {\bf $ f(R, T) $ gravity \cite{Nojiri:2017ncd, dosSantos:2018nmu, Elizalde:2018arz, Singh:2024kez, Singh:2023gxd, Sharif:2019hyl, Bhattacharjee:2020eec, Singh:2022jue, Singh:2022eun, Nagpal:2018mpv, Singh:2018xjv, Singh:2024ckh, Shaily:2024nmy, Harko:2011kv, Nojiri:2010wj},} $ f(R, G) $ gravity \cite{Shaily:2024rjq, Singh:2024aml, Singh:2022gln, Rani:2024uah}, $ f(R, L_m) $ gravity \cite{Singh:2022ptu, Singh:2024gtz}, $ f(Q, C) $ gravity \cite{Shaily:2024tmx}, and many other works have been discussed in modified theories of gravity \cite{Nojiri:2004bi, Singh:2024zvm, Shaily:2022enj, Capozziello:2023vne, Capozziello:2022zzh, Balhara:2023mgj, Singh:2022wwa, Gohain:2024tnf, Shaily:2024xho, Goswami:2022vfq, Singh:2015hva}. Jean-Luc Lehners studied the ekpyrotic and cyclic phases of the universe, which provides the theory of the very early and very late universes. In these models, the Big Bang is described as a collision of branes and thus the Big Bang is not the beginning of time. Before the big bang, there was an ekpyrotic phase with the equation of state $ \omega \geq 1 $  during which the universe slowly contracts \cite{Lehners:2008vx}.

\qquad Recently, the study of modified theories of gravity has focused on $ f(Q, T) $ gravity. The $ f(Q, T) $ gravity is an extension of $ f(Q) $ gravity that is developed with the help of the symmetric teleparallel equivalent of GR. In $ f(Q) $ gravity, curvature, and torsion vanish, leaving just the non-metricity tensor $ Q $ to describe the gravitational interaction. Several different properties for this gravity have been discussed \cite{Koussour:2022sab, Koussour:2022ycn, Koussour:2022wbi, Koussour:2022irr, Narawade:2022jeg, Goswami:2023knh, Bajardi:2020fxh, Hu:2023ndc, Harko:2018gxr, Capozziello:2024vix, Hu:2023gui, Heisenberg:2023lru, Khyllep:2021pcu, Dimakis:2021gby, Dimakis:2022rkd, Paliathanasis:2023nkb, Yang:2024tkw}. The gravity theory $ f(Q, T) $, that combines the effects of the non-metricity tensor and the trace of the EMT, has been the subject of several studies \cite{Arora:2020met, Myrzakulov:2023ohc, Arora:2021jik, Kale:2023nvx, Najera:2022lkf, Yang:2021fjy, Gadbail:2021kgd, Gadbail:2021fjf, Godani:2021mld, Xu:2019sbp, Arora:2020tuk, Arora:2020iva, Najera:2021afa, Xu:2020yeg, Kaczmarek:2024quk}. The historical development of gravitational theories leading up to $ f(Q, T) $ gravity has been marked by significant advancements. Harko and Lobo \cite{Harko:2018ayt} expanded on the development of modified theories of gravity, including those with generalized curvature-matter couplings and hybrid metric-Palatini gravity, setting the stage for the emergence of $ f(Q, T) $ gravity. The review by Bahamonde et al. \cite{Bahamonde:2021gfp} explores the theoretical foundations and advancements in teleparallel gravity, including its development as a gauge theory, modified formulations like the teleparallel Horndeski analogue, and its applications in astrophysics, cosmology, and precision studies. Golovnev and Guzm\'an \cite{Golovnev:2020zpv} discussed foundational issues in $ f(T) $ gravity theory, which shares similarities with $ f(Q, T) $ gravity and provides valuable insights into the theoretical underpinnings of these modified gravitational theories. There is a sub-classification of $ f(Q, T) $ gravity known as Weyl type $ f(Q, T) $ gravity in which the function $ f(Q, T) $ is selected to adhere to specific properties consistent with the Weyl geometry \cite{Yang:2021fjy, Gadbail:2021kgd, Gadbail:2021fjf}.    

\qquad The concept of $ f(Q, T) $ gravity, an extension of symmetric teleparallel gravity, has significant implications in theoretical physics and various astrophysical and cosmological contexts. It introduces a function $ f $ of the non-metricity tensor $ Q $ and the trace of the matter EMT $ T $, leading to non-conservation of energy-momentum and accelerating expansion of the universe \cite{Xu:2019sbp, Arora:2020iva}. This theory has been applied to various cosmological models with and without dark energy or dark matter. It has been found to obey energy conditions and support the accelerating behavior of the universe \cite{Heisenberg:2023lru, Xu:2020yeg}. These findings suggest that $ f(Q, T) $ gravity has the potential to reshape our understanding of gravity and its manifestations in cosmology. Cai et al. \cite{Cai:2015emx} and Kofinas, along with Saridakis \cite{Kofinas:2014daa}, both explore its cosmological applications, with Cai focusing on the late-time universe acceleration and Kofinas on the unified description of cosmological history. Heisenberg \cite{Heisenberg:2023lru} highlights its potential to provide new insights into the nature of gravity, address computational and conceptual questions, and reshape our understanding of gravity in cosmology. This paper also further discusses its potential in early and late-time cosmology, black holes, and wormholes, without the need for dark energy, inflaton fields, or dark matter. These studies collectively demonstrate the versatility and potential of $ f(Q, T) $ gravity in advancing our understanding of the universe. Godani and Samanta \cite{Godani:2021mld} further investigate its application in the Friedmann-Robertson-Walker (FRW) model, using Hubble and Supernova data to constrain model parameters and compare results with the $ \Lambda $CDM model. The development of cosmological perturbation theory in $ f(Q, T) $ gravity has provided a framework for testing these models with observational data, potentially shedding light on the Hubble constant tension and the nature of dark matter \cite{Najera:2021afa}. 

\qquad The content of the work is organized as follows: Sect. \ref{outline} contains an overview of the $ f(Q, T) $ theory, the process for finding the field equations using the Einstein-Hilbert action of $ f(Q, T) $ gravity, and evaluating various physical parameters. In Sect. \ref{observe}, we find the best-fit values of the model parameters using the Hubble datasets of $ 77 $ points, the $ Pantheon $ datasets of $ 1048 $ points, and their joint datasets to constrain all the model parameters. In Sect. \ref{res}, we analyze the behavior of the various cosmological parameters using the constrained model parameters. The analysis of the energy conditions, the scalar field, and the slow-roll parameters has been performed in the various subsections of the Sect. \ref{res}. Finally, we interpret the physical results of the model in Sect. \ref{concs}.

\section{ The $ f(Q,T) $-gravity and the Einstein field equations}\label{outline}

The Einstein-Hilbert action for the modified gravity theory $ f(Q,T) $ is \cite{Xu:2019sbp}
\begin{equation}\label{1}
	S = \int \left[ \frac{1}{16\pi}f(Q,T) + S_m \right] \sqrt{-g} d^4x,
\end{equation}
where $ f(Q, T) $ is any function depends $ Q $ and $ T $. The notation $ S_m $ is for the Lagrangian of a given matter. The quantity $ Q $ is given by
\begin{equation}\label{2}
	Q \equiv -g^{\mu\nu} \left( {L^\alpha}_{\beta\mu} {L^\beta}_{\nu\alpha} - {L^\alpha}_{\beta\alpha} {L^\beta}_{\mu\nu} \right),
\end{equation}
where $ {L^\alpha}_{\beta\gamma} $, the disformation tensor can be calculated by the definition
\begin{equation}\label{3}
	{L^\alpha}_{\beta\gamma} \equiv -\frac{1}{2}g^{\alpha\lambda} \left( \nabla_\gamma g_{\beta\lambda} + \nabla_\beta g_{\lambda\gamma} - \nabla_\lambda g_{\beta\lambda} \right).
\end{equation}

The variation of the action represented by Eq. (\ref{1}) gives the field equations obtained as
\begin{equation}\label{4}
	-\frac{2}{\sqrt{-g}} \nabla_\alpha \left( f_Q \sqrt{-g} {P^\alpha}_{\mu\nu} \right) - \frac{1}{2} fg_{\mu\nu} + f_T \left( T_{\mu\nu}+\Theta_{\mu\nu} \right) - f_Q \left( P_{\mu\alpha\beta} {Q_\nu}^{\alpha\beta} - 2{Q^{\alpha\beta}}_\nu P_{\alpha\beta\nu} \right) = 8\pi T_{\mu\nu}.
\end{equation}
Here, $ f_Q = \frac{\partial f}{\partial Q} $, $ f_T = \frac{\partial f}{\partial T} $, and $ {P^\alpha}_{\mu\nu} $ is the superpotential of the model \cite{Xu:2019sbp}. The terms energy-momentum tensor ($ T_{\mu\nu} $) and $ \Theta_{\mu\nu} $ are written as
\begin{equation}\label{5}
	T_{\mu\nu} \equiv -\frac{2}{\sqrt{-g}}\frac{\delta \left(\sqrt{-g}S_m \right)}{\delta g^{\mu\nu}}
\end{equation}
and
\begin{equation}\label{6}
	\Theta_{\mu\nu} \equiv g^{\alpha\beta} \frac{\delta T_{\alpha\beta}}{\delta g^{\mu\nu}}. 
\end{equation}

The flat FLRW (Friedmann-Lemaître-Robertson-Walker) metric is defined by the line element as
\begin{equation}\label{7}
    ds^2 = -dt^2 + a^2(t)\sum_{i=1}^3 dx_i^2,
\end{equation}
being $ a(t) $ the scale factor. We consider a non-metricity scalar $ Q=6H^2 $, and the energy-momentum tensor for the perfect fluid-filled universe is taken as
\begin{equation}\label{8}
	T^\mu_\nu = diag(-\rho,p,p,p),
\end{equation}
where $ \rho $ and $ p $ are the energy density and isotropic pressure of the universe, respectively. Using Eq. (\ref{8}), $ T=-\rho+3p $. The expression for the tensor $ \Theta^\mu_\nu $ is given by
\begin{equation}\label{9}
	\Theta^\mu_\nu = diag(2\rho+p,-p,-p,-p).
\end{equation}

Using Eqs. (\ref{4}), (\ref{5}), (\ref{6}), and (\ref{7}), the generalized Friedmann equations can be determined as
\begin{equation}\label{10}
	8\pi\rho = \frac{f}{2} - 6FH^2 - \frac{2\tilde{G}}{1+\tilde{G}}\left( \dot{F}H+F\dot{H} \right),
\end{equation}
\begin{equation}\label{11}
	8\pi p = -\frac{f}{2} + 6FH^2 + 2\left( \dot{F}H+F\dot{H} \right),
\end{equation}
where $ F \equiv f_Q $, $ 8\pi\tilde{G} \equiv f_T $, and the overhead dot represents the differentiation of that respective function concerning time $ t $. The generalized form of $ f(Q, T) $ gravity provides more flexibility in exploring cosmological models. Thus, we take the function $ f(Q, T)=\zeta Q^n + \gamma T^m $, $ \zeta<0 $.


Using Eqs. (\ref{10}), (\ref{11})and $ f(Q, T) $, the field equations yield
\begin{equation}\label{12}
8\pi\rho = \frac{1}{2} \left(\zeta Q^n + \gamma T^m\right) - 6n\zeta Q^{n-1} H^2 - \frac{2m\gamma T^{m-1}}{8\pi + m\gamma T^{m-1}} \left( 2n(n-1) (6)^{n-1} \zeta H^{2n-2} \dot{H} + n\zeta Q^{n-1} \dot{H} \right),
\end{equation}
\begin{equation}\label{13}
8\pi p = -\frac{1}{2} \left(\zeta Q^n + \gamma T^m\right) + 6n\zeta Q^{n-1} H^2 + 2\left( 2n(n-1) (6)^{n-1} \zeta H^{2n-2} \dot{H} + n\zeta Q^{n-1}\dot{H} \right).
\end{equation}
The field equations (\ref{12}) and (\ref{13}) are highly non-linear and have many unknowns, so to find observationally consistent solutions, considering the suitable values of $ n $ and $ m $ becomes a necessity. Since we still want to study the non-linear contributions of $ Q $ or $ T $, we take the values $ n=2 $ and $ m=1 $, which is motivated by some distinct research works \cite{Xu:2019sbp, Arora:2020tuk}.


Now, taking $ Q=6H^2 $, $ T=-\rho+3p $, $ n=2 $, and $ m=1 $, the field equations (\ref{12}), (\ref{13}) reduce to
\begin{equation}\label{14}
\left(\frac{16\pi+\gamma}{2}\right)\rho = \frac{3\gamma}{2}p - 54\zeta H^4 - \frac{72\gamma\zeta}{8\pi+\gamma}H^2\dot{H},
\end{equation}
\begin{equation}\label{15}
\left(\frac{16\pi+3\gamma}{2}\right)p = \frac{\gamma}{2}\rho + 54\zeta H^4 + 72\zeta H^2\dot{H}.
\end{equation}

Solving Eqs. (\ref{14}) and (\ref{15}), we get the energy density $ \rho $ and isotropic pressure $ p $ as
\begin{equation}\label{16}
\rho = \frac{9\zeta H^2 \left( -3(8\pi+\gamma) H^2 + 2\gamma \dot{H} \right)}{(4\pi+\gamma)(8\pi+\gamma)},
\end{equation}
and
\begin{equation}\label{17}
p = \frac{9\zeta H^2 \left( 3(8\pi+\gamma) H^2 + 2 (16\pi+3\gamma) \dot{H} \right)}{(4\pi+\gamma)(8\pi+\gamma)}.
\end{equation}

We consider the jerk parameter $ j $ as a function of the redshift $ z $, and is given by
\begin{equation}\label{18}
j(z) = q(z)+2q(z)^2+(1+z)\frac{dq(z)}{dz}.
\end{equation}

Eq. (\ref{18}) contains two unknown parameters $ j(z) $ and $ q(z) $. So, we assume one of these two parameters to evaluate the other. Some of the researchers parametrized the jerk parameter to construct their cosmological models \cite{Rapetti:2006fv, Zhai:2013fxa}. Similarly, we consider $ j(z)=e^{q(z)} $ and tried to solve the differential equation (\ref{18}). To reduce the intricacy of the calculations, we take the first three terms of Maclaurin's expansion of $ e^{q(z)} $ as $ j(z)=1+q(z)+\frac{q^2(z)}{2} $.

Now, putting $ j(z) = 1+q(z)+\frac{q^2(z)}{2} $ in Eq. (\ref{18}) and solving, we get
\begin{equation}\label{19}
q(z) = -\frac{\sqrt{2/3} \left( e^{2\sqrt{6}\alpha} - \left(1+z\right)^{\sqrt{6}} \right)}{e^{2\sqrt{6}\alpha} + \left(1+z\right)^{\sqrt{6}}},
\end{equation}
where $ \alpha $ is an arbitrary integration constant.

The deceleration parameter $ q $ in terms of the Hubble parameter $ H $ and redshift $ z $ is 
\begin{equation}\label{20}
q(z) = -1 +\frac{(1+z)}{H(z)}\frac{dH}{dz}.
\end{equation}

Using Eqs. (\ref{19}) and (\ref{20}), the Hubble parameter $ H(z) $ is given by
\begin{equation}\label{21}
H(z) = \beta \left(1+z\right)^{1-\sqrt{2/3}} \left( e^{2\sqrt{6}\alpha} + \left(1+z\right)^{\sqrt{6}} \right)^{2/3},
\end{equation}
where $ \beta $ is an arbitrary integration constant.

Here, we define $ \beta $ as $ \frac{H_0}{\left( e^{2\sqrt{6}\alpha}+1 \right)^{2/3}} $, where $ H_0 $ is the present value of the Hubble parameter $ H(z) $, and Eq. (\ref{21}) reduce to

\begin{equation}\label{22}
H(z) = \frac{H_0}{\left( e^{2\sqrt{6}\alpha}+1 \right)^{2/3}} \left(1+z\right)^{1-\sqrt{2/3}} \left( e^{2\sqrt{6}\alpha} + \left(1+z\right)^{\sqrt{6}} \right)^{2/3}.
\end{equation}

\section{ The Statistical analysis of the model}\label{observe}

To estimate the observational constraints of the model parameters, we use different datasets employed in this work and deliberate about the methodology used to constrain the free parameters. The Markov Chain Monte Carlo (MCMC) sampling technique optimizes the values of model parameters in Python coding with the EMCEE library. We employ the newly published $ H(z) $ dataset and the $ Pantheon $ dataset in Python coding. The $ H(z) $ dataset contains $ 77 $ observations of cosmic chronometers that span a redshift range $ 0\leq z \leq 2.36 $. The $ Pantheon $ dataset comprises $ 1048 $ observations of Supernova Type Ia that span a redshift range $ 0.01 \leq z \leq 2.26 $. The $ Pantheon $ datasets have been collected from various surveys, namely, high-$ z $, PS1, SDSS, SNLS, low-$ z $, HST, CfA 1-4, etc. (refer to Table \ref{tabpan}) \cite{Asvesta:2022fts}.

\begin{table}[H]
\caption{ The $ Pantheon $ datasets survey.}
\begin{center}
\label{tabpan}
\begin{tabular}{l c c c} 
\hline\hline
\\ 
{Surveys} &      ~~~~~  The data points & ~~~~~ Redshift range &   ~~~~ References
\\
\\
\hline      
\\
{CfA 1-4 }     & ~~~~~ $ 147 $   &  ~~~~~ $ 0.01-0.07 $  &  ~~~~~~~~ \cite{Riess:1998dv, Jha:2005jg, Hicken:2009df, Hicken:2009dk, Hicken:2012zr}
\\
\\
{CSP }     &  ~~~~~ $ 25 $   &  ~~~~~ $ 0.01-0.06 $  &  ~~~~~~~~ \cite{Contreras:2009nt}
\\
\\
{SDSS }  & ~~~~~ $ 335 $   &  ~~~~~ $ 0.03-0.40 $  &  ~~~~~~~~ \cite{SDSS:2014irn}
\\
\\
{SNLS }  & ~~~~~ $ 236 $   &  ~~~~~ $ 0.12-1.06 $  &  ~~~~~~~~ \cite{SNLS:2010pgl}
\\
\\
{PS1 }  & ~~~~~ $ 279 $   &  ~~~~~ $ 0.02-0.63 $  &  ~~~~~~~~ \cite{Pan-STARRS1:2017jku}
\\
\\
{high-$z$ }  & ~~~~~ $ 26 $   &  ~~~~~ $ 0.73-2.26 $  &  ~~~~~~~~ \cite{SupernovaCosmologyProject:2011ycw, Riess:2017lxs, SupernovaSearchTeam:2004lze, Riess:2006fw}
\\
\\
\hline
\\
{Total }  & ~~~~~ $ 1048 $   &  ~~~~~ $ 0.01-2.26 $
\\
\\
\hline\hline  
\end{tabular}    
\end{center}
\end{table}

It is noted that $ H(z)=-\frac{1}{1+z} \frac{dz}{dt} $. Using the $ H(z) $ dataset we calculate the best-fit values of model parameters $ \alpha $ and $ H_0 $ by minimizing the value of $ \chi^2 $. The formula for minimizing $ \chi^2 $ is given by

\begin{equation}\label{23}
    \chi _{HDS}^{2}(\alpha, H_0)=\sum\limits_{i=1}^{77}\left[ \frac{H_{th}(\alpha,H_0,z_{i})-H_{ob}(z_{i})}{\sigma _{H(z_i)}}\right] ^2,
\end{equation}
where $ \chi _{HDS}^{2} $ represents the value of $ \chi^2 $ for the Hubble dataset, $ H(z_i) $ is obtained at redshift $ z_{i} $, and $ H_{th} $ and $ H_{ob}$ indicate the theoretical and observed values of Hubble parameter, respectively. The standard error in $ H_{ob} $ is indicated by $ \sigma_{H(z_{i})} $. 

\begin{figure}\centering
	\subfloat[]{\label{a}\includegraphics[scale=0.75]{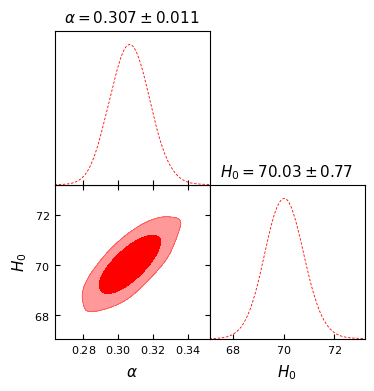}}\hfill
	\subfloat[]{\label{b}\includegraphics[scale=0.75]{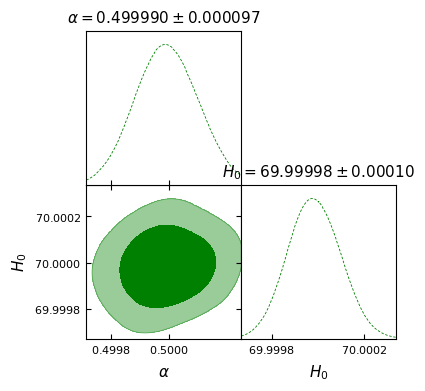}}\par
	\subfloat[]{\label{c}\includegraphics[scale=0.75]{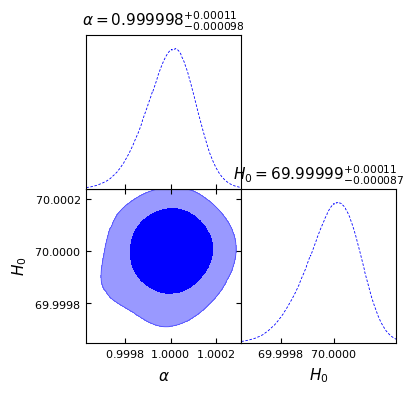}}
\caption{ The posterior distribution of the $ f(Q, T) $ model for $ 1\sigma $ and $ 2\sigma $ CLs obtained from the OHD, $ Pantheon $, and OHD+$ Pantheon $ datasets respectively.}
\label{contours}
\end{figure}

The $ Pantheon $ data collected through various surveys plays a prominent role in investigating the expanding nature of the universe. Using the theoretically estimated apparent magnitude ($ m $) and absolute magnitude ($ \mathcal{M} $) in regards to the color and the stretch, we determine the value of the distance modulus $ \mu_{th}(z_i) $ as
\begin{equation}\label{24}
\mu(z)= m-\mathcal{M} = 5Log \mathcal{D}_L(z)+\mu_{0}.
\end{equation}

For the flat universe, the luminosity distance $ \mathcal{D}_L(z)=(1+z) c \mathcal{D}_m \int_0^z \frac{1}{H(w)}dw $, where $\mathcal{D}_m(z)$ is defined as

\begin{equation}\label{25}
\mathcal{D}_m(z)=\begin{cases}
\frac{\sinh(\sqrt{\Omega_n})}{H_0 \sqrt{\Omega_n}}, ~~~~\text{for} ~~\Omega_n>0\\
1, ~~~~~~~ \text{for} ~~ \Omega_n=0\\
\frac{\sin(\sqrt{\Omega_n})}{H_0 \sqrt{\Omega_n}}, ~~~~ \text{for} ~~\Omega_n<0
\end{cases} 
\end{equation}

and the nuisance parameter $ \mu_0 $ is given as

\begin{equation}\label{26}
\mu_0= 5Log\Big(\frac{H_0^{-1}}{1Mpc}\Big)+25.
\end{equation}

The maximum likelihood method is used to constrain the values of $\alpha$ and $H_0$. The relevant $ \chi^2 $ function is given by

\begin{equation}\label{27}
\chi _{PDS}^{2}(\alpha,H_0)=\sum\limits_{i=1}^{1048}\left[ \frac{\mu_{th}(\alpha,H_0,z_{i})-\mu_{obs}(z_{i})}{\sigma _{\mu(z_{i})}}\right] ^2,
\end{equation}

where PDS represents the $ Pantheon $ dataset. $ \mu_{th} $ and $ \mu_{obs} $ are used for the theoretical and observed distance moduli of this model, respectively, and $ \sigma_\mu (z_i) $ indicates the standard error in the observed value.

We employ joint statistical analysis using both datasets, that is, Hubble samples and $ Pantheon $ samples. This joint analysis gives better constraints. For this purpose, the $ \chi^2 $ function is specified as
\begin{equation}\label{28}
\chi_{Joint}^{2}=\chi_{OHD}^{2}+\chi_{PDS}^{2}.
\end{equation}

For all three datasets, the constrained values of the model parameters have been summarized in Table \ref{tabparm}.

\begin{table}[htbp]
\caption{The constrained values of the model parameters.}
\begin{center}
\label{tabparm}
\begin{tabular}{l c c c c c r} 
\hline\hline
\\ 
{Dataset} &      ~~~~~  $ \alpha $ & ~~~~~ $ H_0 $  & ~~~~~ $ z_{tr} $\footnote{Transitions from deceleration to acceleration} & ~~~~~ $ q_0 $\footnote{The present value of the $ q $}& ~~~~~ $ p_{tr} $\footnote{Transitions of isotropic pressure from positive to negative}\footnote{$\lim_{z\to -1}p =0 $}
\\
\\
\hline      
\\
{OHD  }     & ~~~~~ $ 0.307 \pm 0.011 $   &  ~~~~~ $ 70.03 \pm 0.77 $ &  ~~~~~ $  0.84781 $ &  ~~~~~ $ -0.519567 $ &  ~~~~~ $ 0.47 $
\\
\\
{ Pantheon  }     &  ~~~~~ $ 0.49999 \pm 0.000097 $   &  ~~~~~ $ 69.99998 \pm 0.0001 $ &  ~~~~~ $ 1.71823 $&  ~~~~~ $  -0.686707 $ &  ~~~~~ $ 1.14 $
\\
\\
{ H(z)+Pantheon}  & ~~~~~ $ 0.99999^{+0.00011}_{-0.000098} $   &  ~~~~~ $ 69.99999^{+0.00011}_{-0.000087} $ &  ~~~~~ $ 6.38903 $&  ~~~~~ $ -0.804414 $ &  ~~~~~ $ 4.79 $
\\
\\ 
\hline\hline  
\end{tabular}    
\end{center}
\end{table}

\begin{figure}\centering
	\subfloat[]{\label{ae}\includegraphics[scale=0.513]{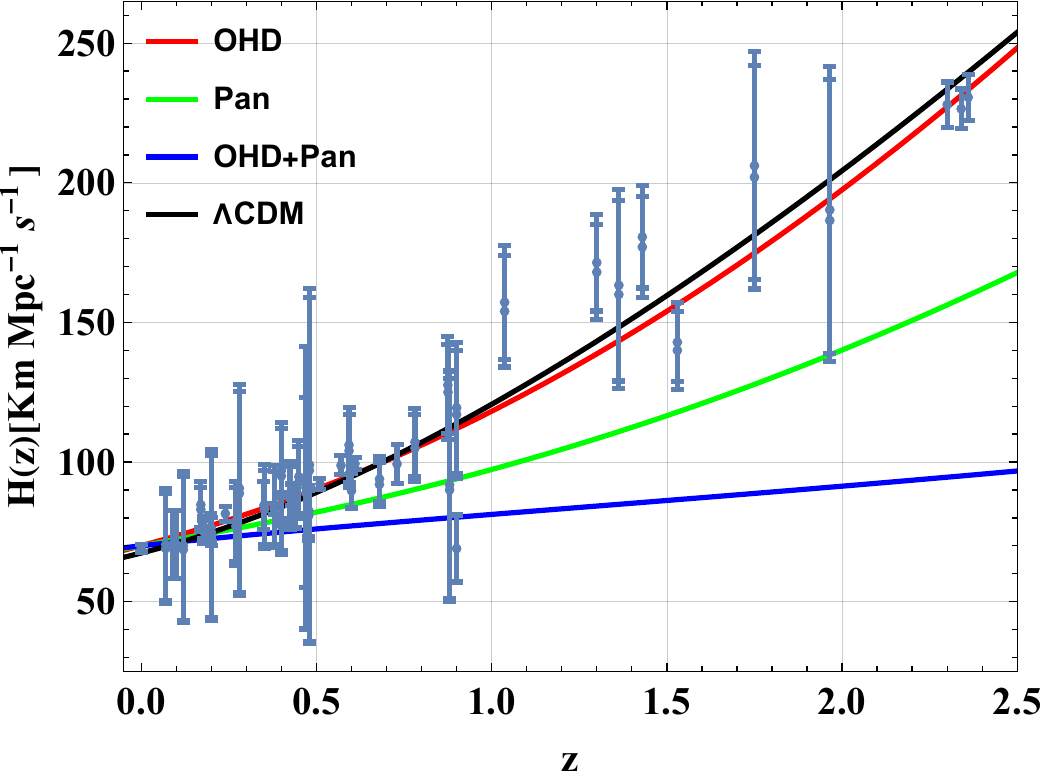}}\hfill
	\subfloat[]{\label{be}\includegraphics[scale=0.49]{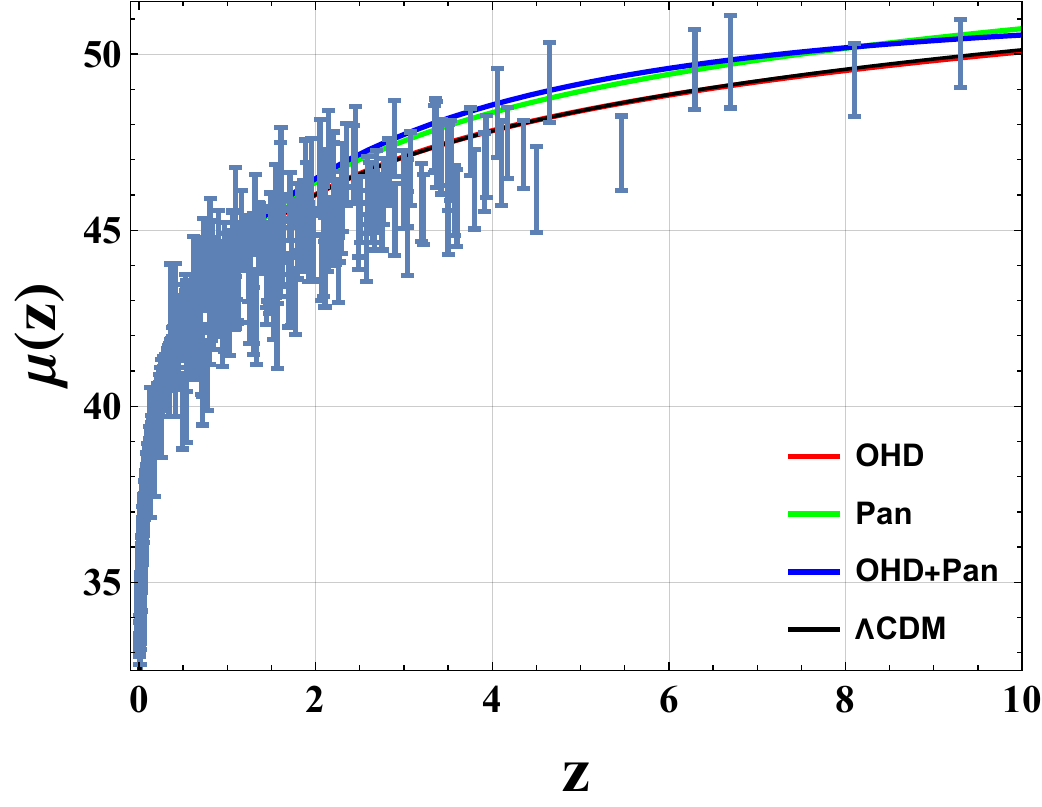}} 
	\caption{The alikeness of our model with $ \Lambda $CDM for OHD, $ Pantheon $, OHD + $ Pantheon $ datasets in the Error Bar plots.}
\label{errorp}
\end{figure}

\section{ The cosmological parameters}\label{res}

In this section, we discuss the physical behaviors of the model. With the help of the scale factor's first-order derivative and higher derivative components, we can analyze the geometrical and dynamic nature of the cosmos more competently. The Hubble parameter $ H $, the deceleration parameter $ q $, and the jerk parameter $ j $ can be expressed as $ H=\frac{\Dot{a}}{a} $, $ q=-\frac{\Ddot{a}a}{\Dot{a}^2} $, $ j=\frac{\dddot{a}a^2}{\Dot{a}^3} $. 

\begin{figure}\centering
	\subfloat[]{\label{bq}\includegraphics[scale=0.5]{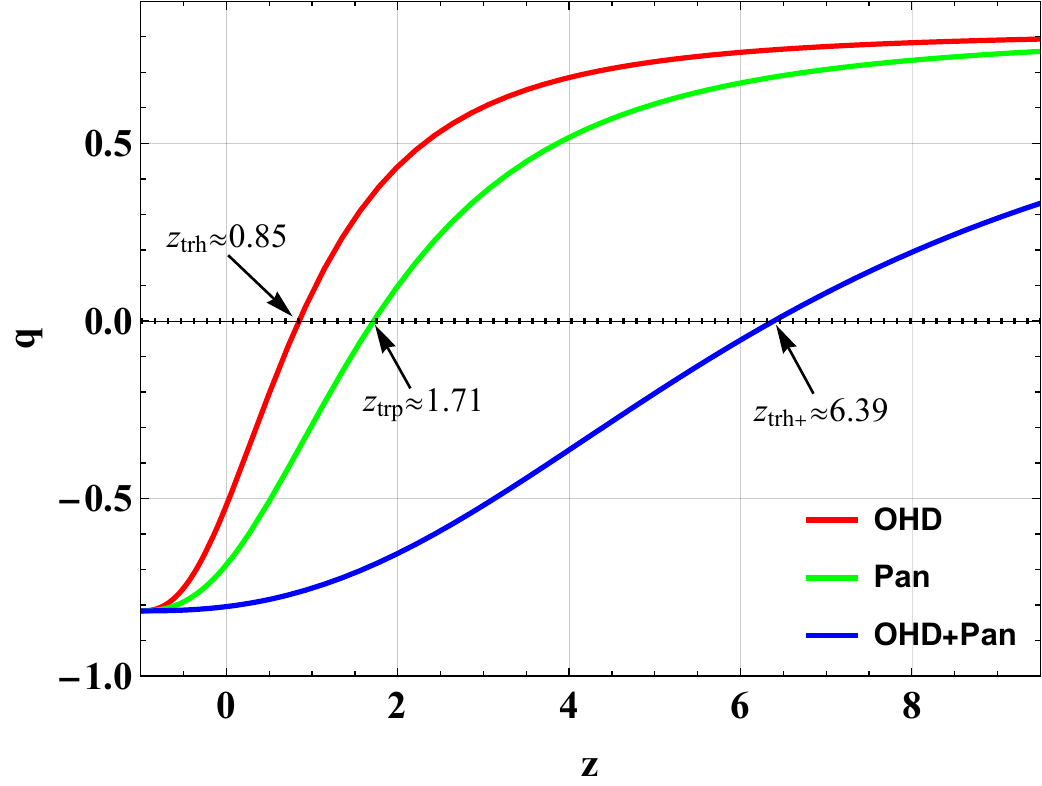}}\hfill 
	\subfloat[]{\label{cj}\includegraphics[scale=0.5]{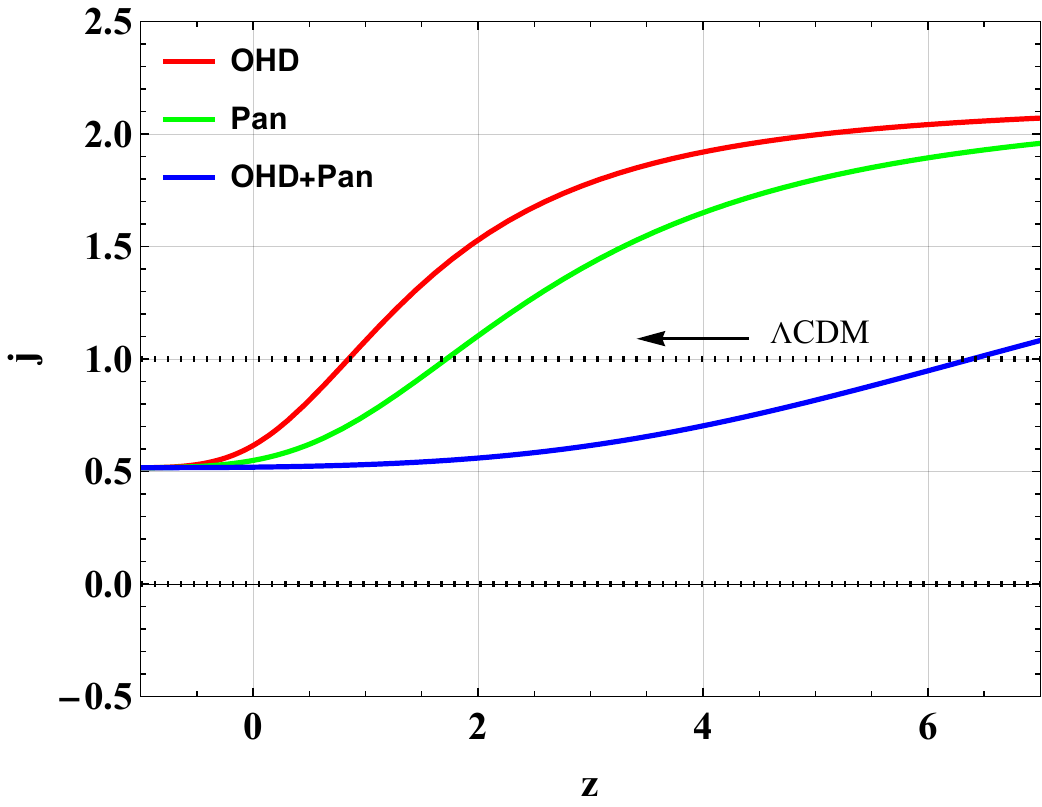}}
	\caption{The evolution of $ q $ and $ j $.}
	\label{Hqjp}
\end{figure}

We analyze the evolution of these cosmic parameters using different plots according to the constraints on $ \alpha $ and $ H_0 $ mentioned in Table \ref{tabparm}. In Fig. \ref{ae}, one can notice that $ H $ decreases monotonically as $ z\to-1 $, showing that the expansion rate is high in the early stages of evolution and decreases in later times. The dimensionless deceleration parameter $ q $ exhibits the transition from early deceleration to acceleration at later times(see Fig. \ref{bq}). The redshift transitions ($ z_{tr} $) are $ 0.847808 $,  $ 1.71823 $, and $ 6.38903 $, and $ q_0 $ are $ -0.519567 $, $ -0.686707 $, and $ -0.804414 $ for different observations shown in Table \ref{tabparm}. Using Eq. (\ref{19}) in the expression $ 1+q(z)+\frac{q^2(z)}{2} $, the jerk parameter $ j $ can be calculated as
\begin{equation}\label{29}
    j=\frac{\left(4-\sqrt{6}\right) e^{4\sqrt{6}\alpha} + 4 e^{2\sqrt{6}\alpha} \left(z+1\right)^{\sqrt{6}} + \left(4+\sqrt{6}\right) \left(z+1\right)^{2\sqrt{6}}}{3 \left( e^{2\sqrt{6}\alpha} + \left(z+1\right)^{\sqrt{6}} \right)^2}.
\end{equation}
In Fig. \ref{cj}, the jerk parameter $ j $ is different from $ 1 $ for each dataset, which indicates that our obtained model deviates from the $ \Lambda $CDM defined by $ j=1 $.

\begin{figure}\centering
	\subfloat[]{\label{ar}\includegraphics[scale=0.51]{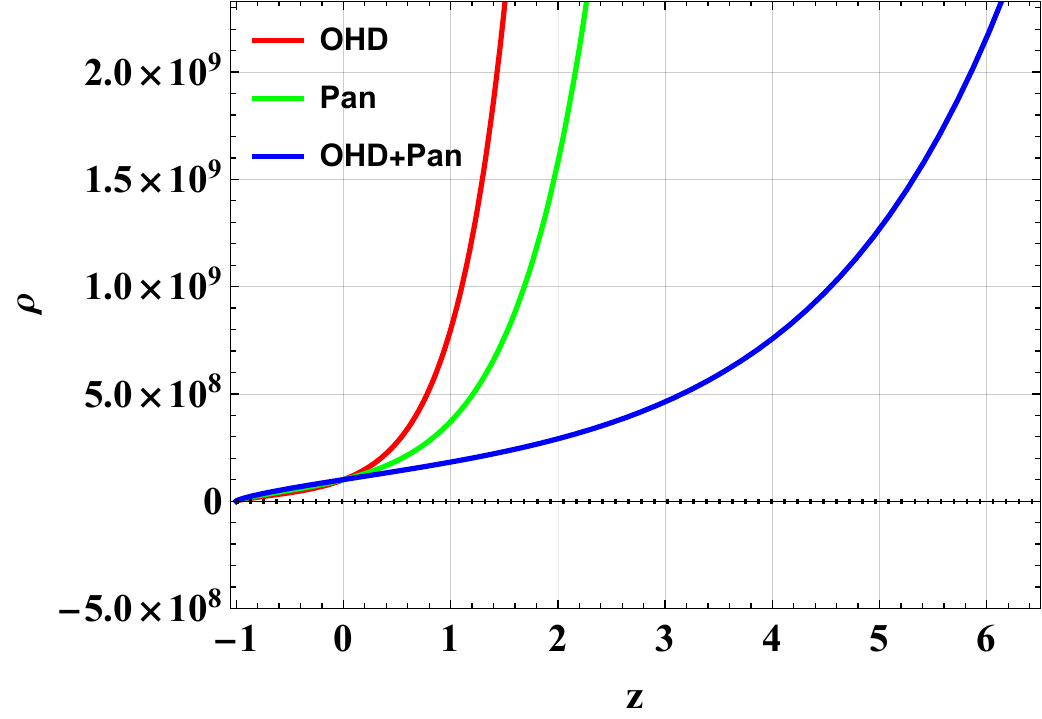}}\hfill
	\subfloat[]{\label{bp}\includegraphics[scale=0.5]{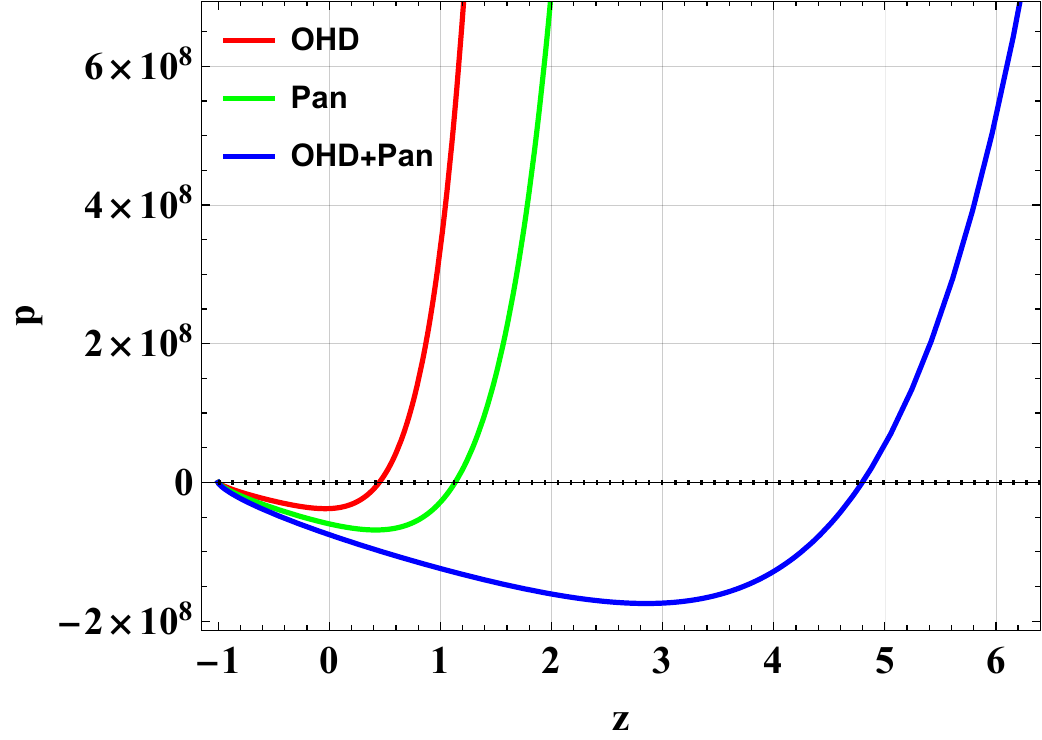}}\par 
	\subfloat[]{\label{cw}\includegraphics[scale=0.5]{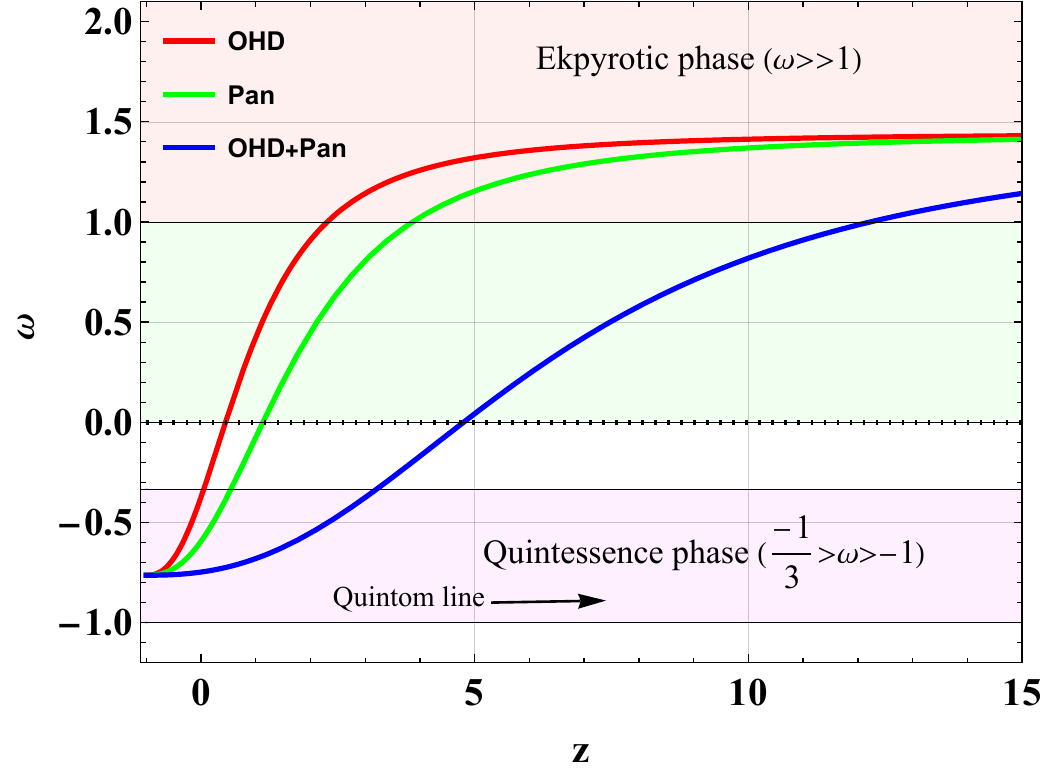}}
	\caption{ The evolution of the $ \rho $, $ p $, and the EoS using $ \zeta=-1.82 $ and $ \gamma=-0.9 $.}
	\label{rhopw}
\end{figure}

\qquad Using the expression $ \dot{H}=-(1+z) H(z) \frac{dH}{dz} $ and Eq. (\ref{22}) in Eqs. (\ref{16}) and (\ref{17}), we evaluate $ \rho $ and p as the functions of $ z $. The trajectories in Figs. \ref{ar} and \ref{bp} depict that $ \rho $ and $ p $ decrease monotonically during the evolution for OHD, $ Pantheon $, and OHD+$ Pantheon $ datasets. In Fig. \ref{bp}, $ p $ transits from a positive to a negative value in the redshift range $ 0.47 \leq z \leq 4.79 $ which shows that our model is accelerated expanding (see Table \ref{tabparm}) \cite{Singh:2022nfm, Singh:2023bjx, Singh:2024urv}. In Fig. \ref{cw}, the trajectories of $ \omega $ pass via various stages of the evolution from very high redshift to low redshift. The EoS parameter is explained as follows:

\begin{table}
\caption{ Description of the substances that exist in the model}
\begin{center}
\label{tabparm2}
\begin{tabular}{l c c r} 
\hline\hline
\\ 
{\textbf{Substance}} & \,\,\,\,\, \textbf{EoS parameter} \,\,\,  &  \,\, \, \textbf{Observations}
\\ 
\\
\hline 
\\
\\
Ekpyrotic matter\footnote{The model provides the ekpyrotic phase of the universe on the high redshift $ z>12.32 $. In this model, the Big Bang is not the initial state of the Universe because the Big Bang is described as a collision of branes. Before the Big Bang, there is an ekpyrotic phase with $ \omega >> 1 $} & $ \omega >> 1 $ & Resist Dominant Energy Condition.
\\
\\
the Stiff-matter filled universe\footnote{leading to a singularity in the early universe.} & $ \omega = 1 $ & Einstein de Sitter (matter dominated) 
\\
\\
Hard Universe & $ \frac{1}{3} < \omega < 1 $ &  Excessive high densities
\\
\\
Radiation era universe\footnote{ The period from about $ 10^{-43} $ Second (the Planck era) to $ 3\times10^4 $ years after the Big Bang. During this time, the expansion of the Universe was dominated by the effects of radiation or high-speed particles (at high energies, all particles behave like radiation).} & $ \omega=\frac{1}{3} $ & Influential in past
\\
\\
Hot matter\footnote{ a scenario where the equation of state parameter, $\omega $ for hot matter lies between $ 0 $ and $ \frac{1}{3} $, which is a region where the matter is considered to be ``hot'' or relativistic.} & $ 0<\omega <\frac{1}{3} $ & Insignificant at present
\\
\\
Pressureless cold dark matter (CDM) & $\omega=0$ & 27\% of the Universe
\\
\\
Quintessence\footnote{ our model shows a quintessence model in the redshift range $ 0.05 \leq z \leq 3.27 $ for all observational datasets OHD, $ Pantheon $, and OHD+$ Pantheon $. } & $ -1< \omega < -\frac{1}{3} $ & 68\% of the Universe
\\
\\ 
Cosmological constant ($ \Lambda $CDM) & $ \omega=-1 $ &  Inconsistent with observations 
\\
\\ 
Phantom Universe & $\omega<-1$ & \,\,\,\,\,Lead to Big Rip, resist Weak Energy Condition
\\
\\ 
\hline\hline  
\end{tabular}    
\end{center}
\end{table}

\subsection{ The energy conditions}\label{ec}

\qquad The energy conditions (ECs) play a significant role in describing space-time singularity problems and analyzing the nature of null, space-like, time-like, or light-like geodesics in classical GR. It provides an extensive range of grounds to study the conduct of cosmic geometries and the correlation that the stress-energy momentum (EMT) $ T_{ij} $ must comply positively. The energy conditions are just primitive constraints on different linear combinations of energy density $ \rho $ and pressure $ p $. It shows that $ \rho >0 $ and gravity are always attractive.

The preliminary ECs discussed in the literature impose certain limitations on the capability of the stress tensor to contract at each location in the space \cite{Curiel:2014zba}. These ECs can also be drafted in a geometric and physical form that complies with Einstein's equation using $ R_{ij} $ and $ T_{ij} $ for many theories. The ECs in various forms are discussed in Table \ref{tab:Energy Conditions}, where $ t^i $ and $ \xi^j $ are coordinated time-like vectors and $ k^i $ is a null (light-like) vector. These ECs are dependent on each other \cite{Kontou:2020bta}. 

\begin{table}[htbp]
\caption{ \textbf{Energy Conditions}}
\centering
\begin{tabular}{l c c c c r}
\hline\hline
\\
Energy Condition & \qquad Physical conformation & \qquad Geometric conformation & \qquad Perfect conformation form
\\
 \\
\hline
\\
 ~~  NEC ~~ & ~~ $ T_{ij}k^ik^j\geq 0 $ & ~~$ R_{ij}k^ik^j\geq 0 $  &  ~~ $ \rho+p \geq 0 $
\\
\\ 
~~  WEC  ~~ & ~~ $ T_{ij}t^it^j\geq 0 $  & ~~$  G_{ij}t^it^j\geq 0 $  & ~~$  \rho \geq 0, \rho+p \geq 0 $  
 \\
\\
~~  SEC  ~~ & ~~  $ (T_{ij}-\frac{T}{n-2}g_{ij})t^it^j\geq 0  $ & ~~ $ R_{ij}t^it^j\geq 0 $  & ~~  $ \rho+p \geq 0, (n-3)\rho+(n-1)p \geq 0 $
\\
\\
~~  DEC ~~  & ~~ $ T_{ij}t^i\xi^j\geq 0 $  & ~~ $ G_{ij}t^i\xi^j\geq 0 $ & ~~   $\rho \geq |p| $  
\\
 \\
\hline\hline
\end{tabular}
\label{tab:Energy Conditions}
\end{table}

\begin{figure}\centering
	\subfloat[]{\label{anec}\includegraphics[scale=0.5]{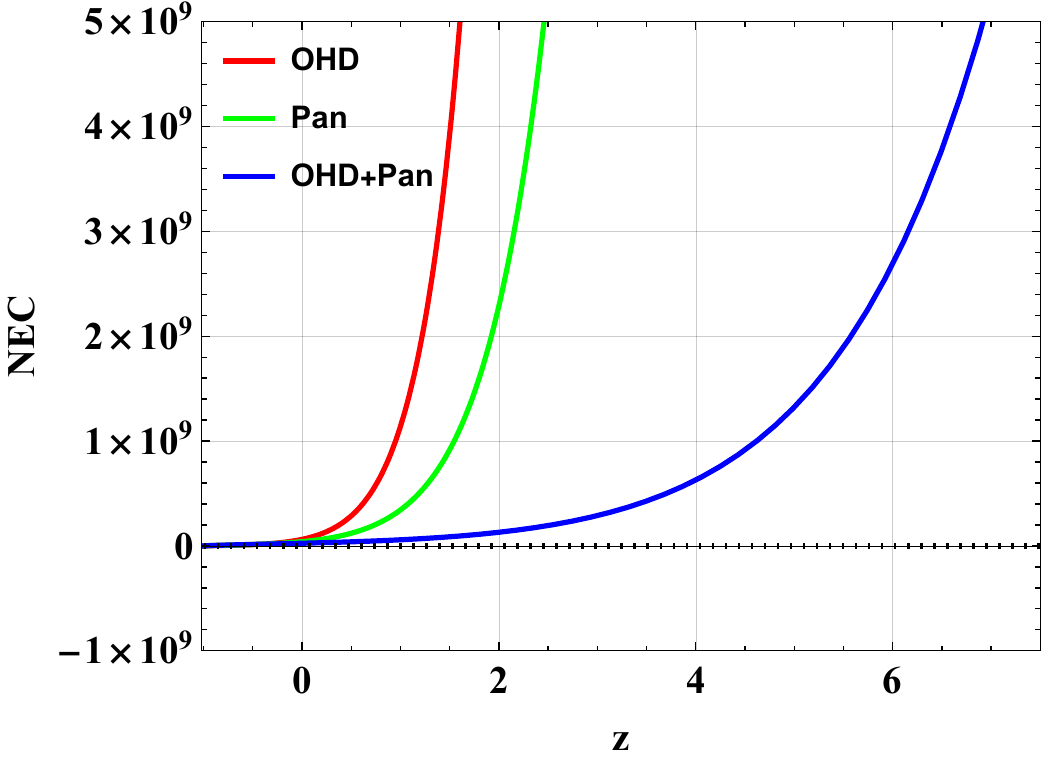}}\hfill
	\subfloat[]{\label{bsec}\includegraphics[scale=0.5]{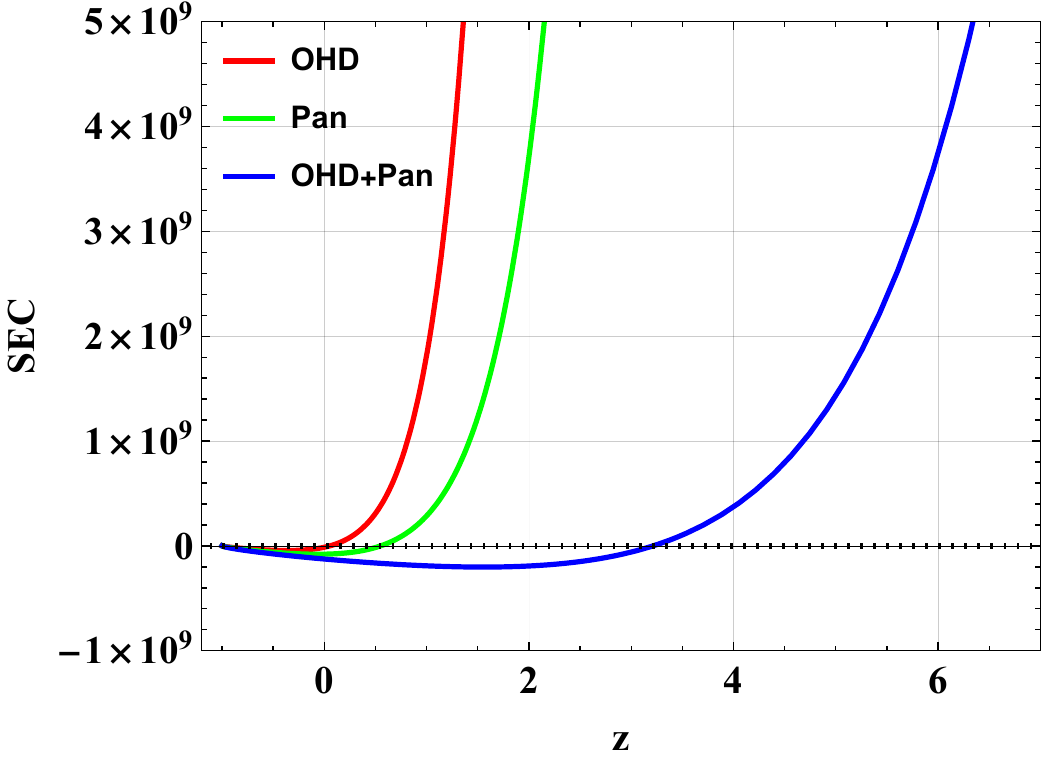}}\par 
	\subfloat[]{\label{cdec}\includegraphics[scale=0.5]{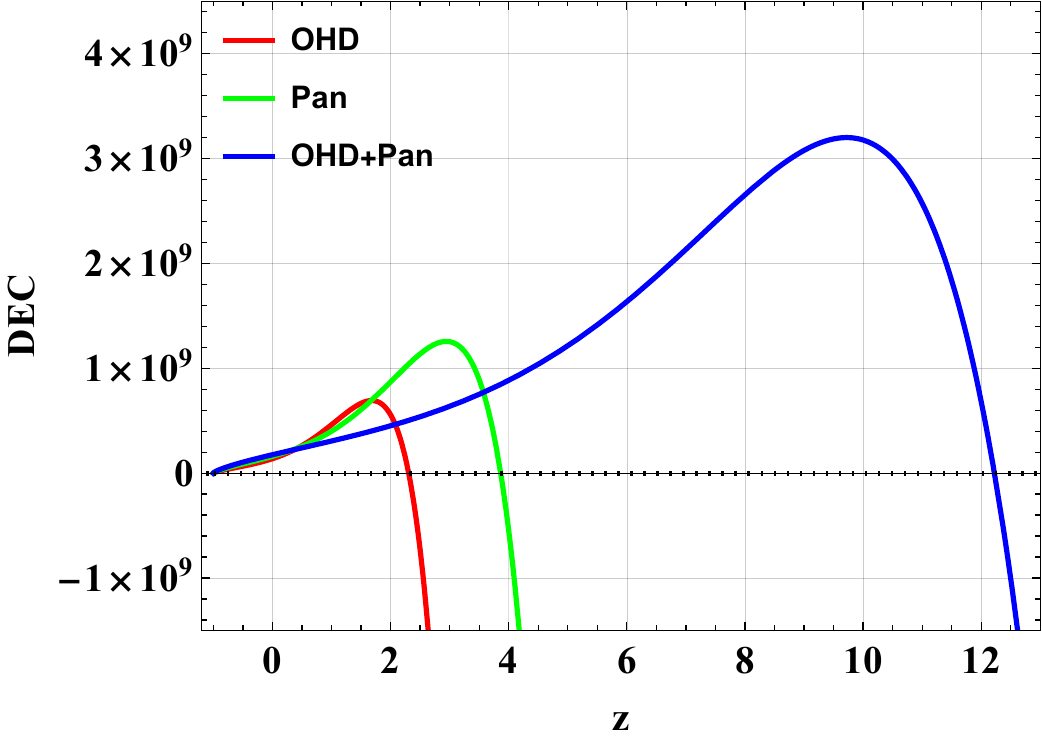}}
	\caption{ The variations of the NEC, SEC, and DEC, respectively.}
	\label{ecs}
\end{figure}
The prior correlations are confined as: \\
(i)   Weak Energy Condition (WEC) $ \Rightarrow $ NEC;\\
(ii)  Strong Energy Condition (SEC) $ \Rightarrow $ NEC; \\ 
(iii) Dominant Energy Condition (DEC) $ \Rightarrow $ NEC.

If the NEC is violated, then all the ECs are violated \cite{Visser:1995cc}. It is trusted that the NEC is valid for all steady, well-behaved models of the Universe. A realistic model can violate other energy criteria by including an appropriate cosmological constant (either positive or negative). Wherever the $ T_{ij} $ violates the NEC, the model comes across destructive uncertainty like a ghost model in which the model has the {\it wrong} sign of the energy or exponentially expanding modes with informally short wavelengths which are referred to as {\it tachyons} \cite{Caldwell:1999ew}. The modified gravity theories can be envisaged as insignificant scalar backgrounds with a $ T_{ij} $ equal to a cosmological constant $ \Lambda $. Consequently, they stand on the extremity of breaking the NEC. This describes how the degenerate dispersion relations came to be, and it also indicates that these theories establish a first step in breaking the NEC.

The NEC plays a pivotal role among several ECs discussed in GR. In this condition, the EMT for matter $ T_{ij} $ satisfies $ T_{ij}k^i k^j $, $ \forall $ $ k^i $ \textit{i.e.}, for any vector satisfying $ g_{ij}k^i k^j=0 $. Exceptionally, the following two bases for the NEC are interesting \cite{Rubakov:2014jja}:\\
(1)  Until rather recently the common mythology was that the NEC could not be violated in a healthy theory, except for scalar field non-minimally coupled to gravity \cite{Flanagan:1996gw}. \\
(2) The NEC is an important expectation of the Penrose singularity theorem \cite{Penrose:1964wq}, sustainable in GR. The theorem expects that (a) the NEC holds; (b) the Cauchy hypersurface is non-compact. The theorem reveals that the future will be singular once a trapped surface exists in space. A trapped surface is a closed surface on which outward-pointing light rays are converging.

The upper left panel of Fig. \ref{ecs} shows that for all datasets, NEC is satisfied in this model. Whereas, Fig. \ref{bsec} depicts that SEC violates our model, and violation of SEC indicates the presence of mysterious substances like dark energy and dark matter. Also, this is consistent with the recent observations. In contrast, the SEC is consistently violated across various best-fit values of the model parameters listed in Table \ref{tabparm}. This violation implies the universe's accelerated expansion in the distant future \cite{Singh:2019fpr, Singh:2024ckh, Singh:2023bjx, Bolotin:2015dja, Visser:1997qk, Visser:1997tq}. 

In Fig. \ref{cdec}, the DEC plot shows that in the early stages, DEC violates, though it satisfies at present and in the later stages.

\subsection{ The scalar field}\label{sf}

\qquad Over the past few years, the quintessence model has convened great recognition in cosmology. This model is characterized by a quintessence-like scalar field because the EoS approaches in the range $ -1 < \omega < -\frac{1}{3} $ which is consistent with recent observations in late times. According to the No-Go theorem, the EoS for a conventional scalar field model, described by a Lagrangian $ L = L(\phi, \partial_\mu \phi \partial^\mu \phi) $, cannot pass over the Quintom line boundary ($ \omega=-1 $) \cite{Xia:2007km, Cai:2007qw}.

The model requisites an EoS close to $ \omega \simeq -1 $ to ensure good agreement with the recent observational datasets. This warrants that the kinetic energy (KE) of the scalar field, $ \dot{\phi}^{2} $, be notably smaller than the potential energy (PE), $ V(\phi) $, i.e., $ \dot{\phi}^{2} << V(\phi) $. When $ \omega \simeq -1 $, several models may be viewed to delineate the universe's acceleration, moreover, it can be redesigned for inflationary reasons. On this ground, our center of attention is on examining the models that utilize the scalar fields enclosed by the structure of $ f(Q, T) $-gravity. This approach enables us to analyze the possible improvement to the standard cosmology and comprehend the implications of modified gravitational theories on the physical features of scalar fields and cosmic acceleration. The EoS $ \omega $ converges to the redshift range $  z \in (-1,-\frac{1}{3}) $, hence this model is a quintessence model (see Fig. \ref{cw}).

The following action is defined in Einstein's theory of GR
\begin{equation}\label{15a}
S=\frac{c^4}{16\pi G}\int R\sqrt{-g}d^{4}x+L_m,
\end{equation}
where $ L_m $ is indicated by $ {L_m}_{q} $, the action for the quintessence-like scalar field and taking $ c=1 $ to normalize.

The action $ {L_m}_{q} $ is given by
\begin{equation}\label{sf1}
	{L_m}_{q} = \int \left( -\frac{1}{2} \partial_\mu \phi_{q} \partial^\mu \phi_{q} - V(\phi_{q}) \right) \sqrt{-g} d^4x.
\end{equation}

\begin{figure}
	\centering 
        \subfloat[]{\label{phiq}\includegraphics[scale=0.5]{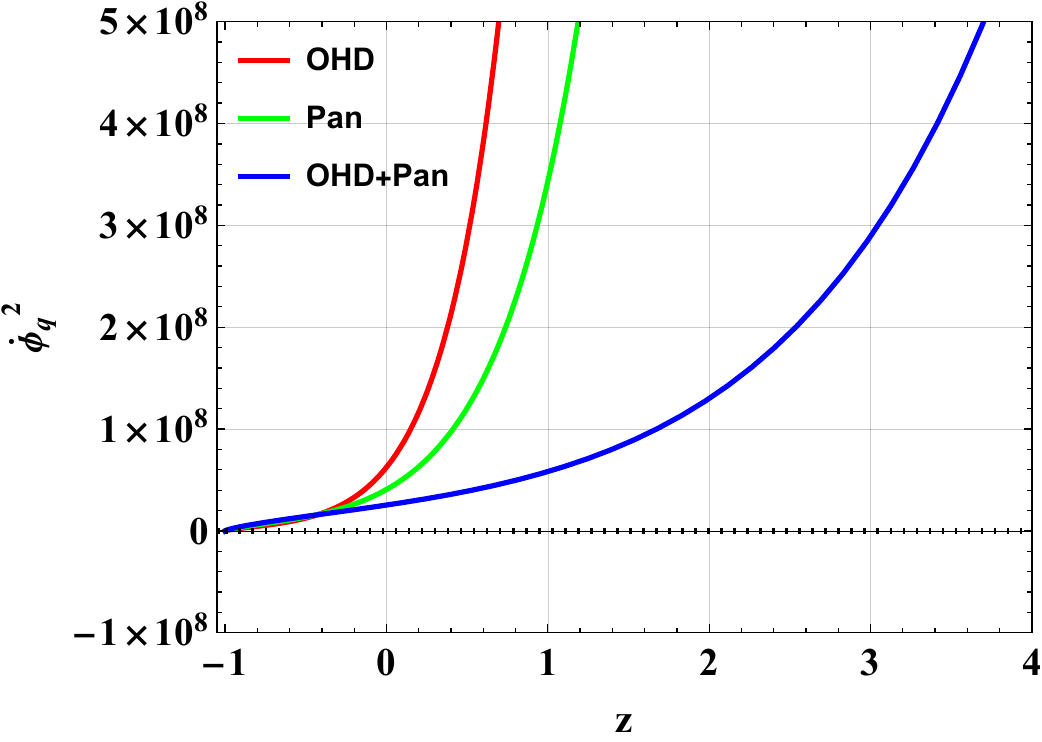}}\hfill
        \subfloat[]{\label{potv}\includegraphics[scale=0.5]{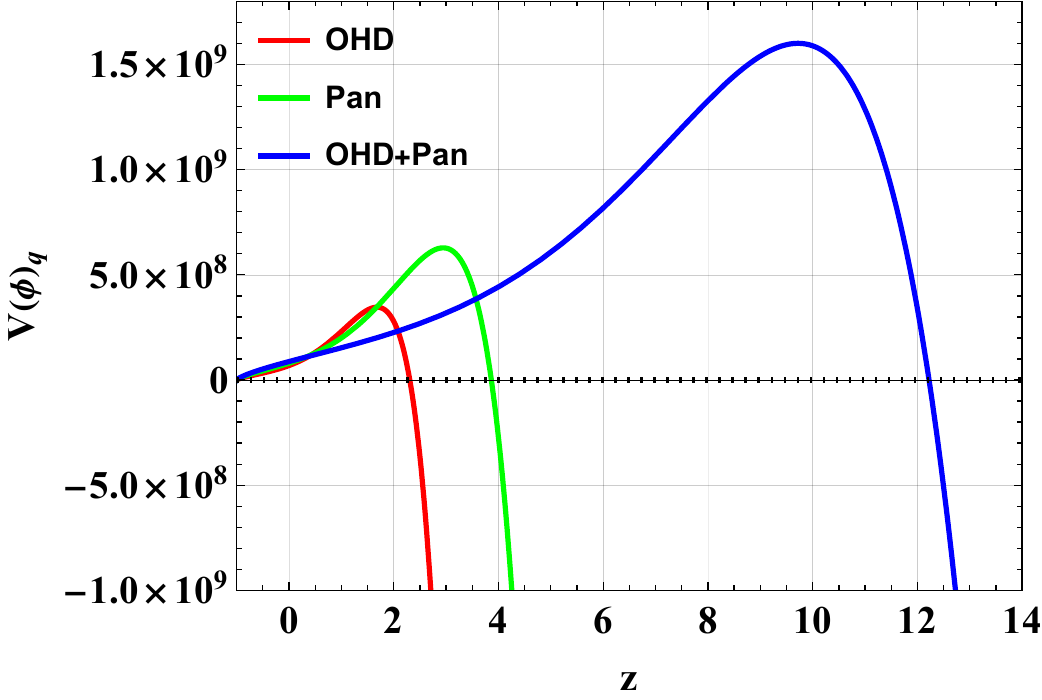}}
		\caption{ The evolution of the $ {{\dot{\phi}}^2}_{q} $ and the $ V(\phi_{q}) $.}
  \label{scalar-z}
  \end{figure}

In this condition, the scalar potential $ V(\phi) $ assimilates to the self-interacting scalar field $ \phi $. Since $ \phi $ is a function of $ t $, therefore, it can be considered like a perfect fluid with energy density $ \rho_\phi $ and pressure $ p_\phi $. Additionally, we can correlate these energy density $ {\rho_\phi}_{q} $ and pressure $ {p_\phi}_{q} $ of the quintessence-like scalar field in the scenario of FLRW cosmology as
\begin{equation}\label{sf3}
	{\rho_\phi}_{q} = \frac{1}{2}{{\dot{\phi}}^2}_{q} + V(\phi_{q}), ~~ {p_\phi}_{q} = \frac{1}{2}{{\dot{\phi}}^2}_{q} - V(\phi_{q}).
\end{equation}

The kinetic energy $ \frac{1}{2}{{\dot{\phi}}^2}_{q} $ and the potential energy $ V(\phi_{q}) $ are given by

\begin{equation}\label{35}
    \frac{1}{2}{{\dot{\phi}}^2}_{q} = -\frac{24 H_0^4 \left(1+z\right)^{4\left(1-\sqrt{2/3}\right)} \left( e^{2\sqrt{6}\alpha} + \left(1+z\right)^{\sqrt{6}} \right)^{5/3} \left(-\left(-3+\sqrt{6}\right) e^{2\sqrt{6}\alpha} + \left(3+\sqrt{6}\right) \left(1+z\right)^{\sqrt{6}} \right) \zeta}{\left(1+e^{2\sqrt{6}\alpha}\right)^{8/3} \left(8\pi+\gamma\right)},
\end{equation}

\begin{equation}\label{37}
    V(\phi_{q}) = -\frac{3 H_0^4 \left(1+z\right)^{4\left(1-\sqrt{2/3}\right)} \left( e^{2\sqrt{6}\alpha} + \left(1+z\right)^{\sqrt{6}} \right)^{5/3} \left( \left(3+2\sqrt{6}\right) e^{2\sqrt{6}\alpha} + \left(3-2\sqrt{6}\right) \left(1+z\right)^{\sqrt{6}} \right) \zeta}{\left(1+e^{2\sqrt{6}\alpha}\right)^{8/3} \left(4\pi+\gamma\right)}.
\end{equation}

In Fig. \ref{scalar-z}, the trajectories describe the evolution of the kinetic term $ \frac{1}{2}{\dot{\phi}^2}_{q} $ and the potential term $ V(\phi_{q}) $ of the energy for the quintessence-like scalar field in terms of the redshift $ z $ in the $ f(Q, T) $ theory of gravity. At the early evolution, the kinetic energy of the quintessence-like field is infinitely large and as time elapses, these values decrease monotonically, approaching zero as $ z \to -1 $ asymptotically.

Initially, the potential energy $ V(\phi_{q}) $ is incapacitated in the redshift range $ z>12.29 $ for the joint dataset, which decreases monotonically over time and tends to zero as $ z \to -1 $ asymptotically in the redshift range less than $ z<12.29 $ (OHD+Pan). Thus, we see that the $ {\dot{\phi}^2}_{q} $ declines  more rapidly than $ V(\phi_{q}) $. The kinetic and potential energies of the quintessence-like field show a similar minimizing nature over time. 

\subsection{ The slow-roll parameters}\label{slroll}

The slow-roll parameter is a fundamental criterion describing the inflationary dynamics, and they have been extensively studied in various cosmological contexts. This parameter can be stated either in terms of the inflaton potential or the Hubble parameter $ H $, allowing for the analysis of both standard and non-standard cosmologies, including the braneworld scenarios \cite{Calcagni:2004bh}. The slow-roll parameter closely resembles the cosmography parameters, which constrain inflationary models and express key inflationary quantities like the spectral index and tensor-to-scalar ratio \cite{Das:2018mog}.

The following constraints must be fulfilled to occur in a new inflationary scenario:
\begin{enumerate}
    \item  The inflaton rolls slowly down the potential, meaning the velocity of the field $ \dot{\phi} $ is small, being $ \phi $ the scalar field corresponding to the potential $ V(\phi) $.
    \item  The curvature of the potential $ V(\phi) $ is sufficiently flat as compared to the large vacuum energy so that $ \ddot{\phi} $ is small.
    \item The inflaton particles must have little mass. 
\end{enumerate}

These conditions are quantified using the slow-roll parameters, the dimensionless quantities that measure how slowly the inflaton field evolves.

The first and the second kind of slow-roll parameters, denoted by $ \epsilon_1 $, and $ \epsilon_2 $ are defined as \cite{Shiravand:2022ccb, Liddle:1994dx}:
\begin{equation}\label{38}
    \epsilon_1(t) = -\frac{\dot{H}}{H^2}.
\end{equation}
and
\begin{equation}\label{39}
    \epsilon_2(t) = \frac{\ddot{H}}{\dot{H} H} - \frac{2\dot{H}}{H^2}.
\end{equation}

\begin{figure}\centering
	\subfloat[]{\label{fsrlp}\includegraphics[scale=0.51]{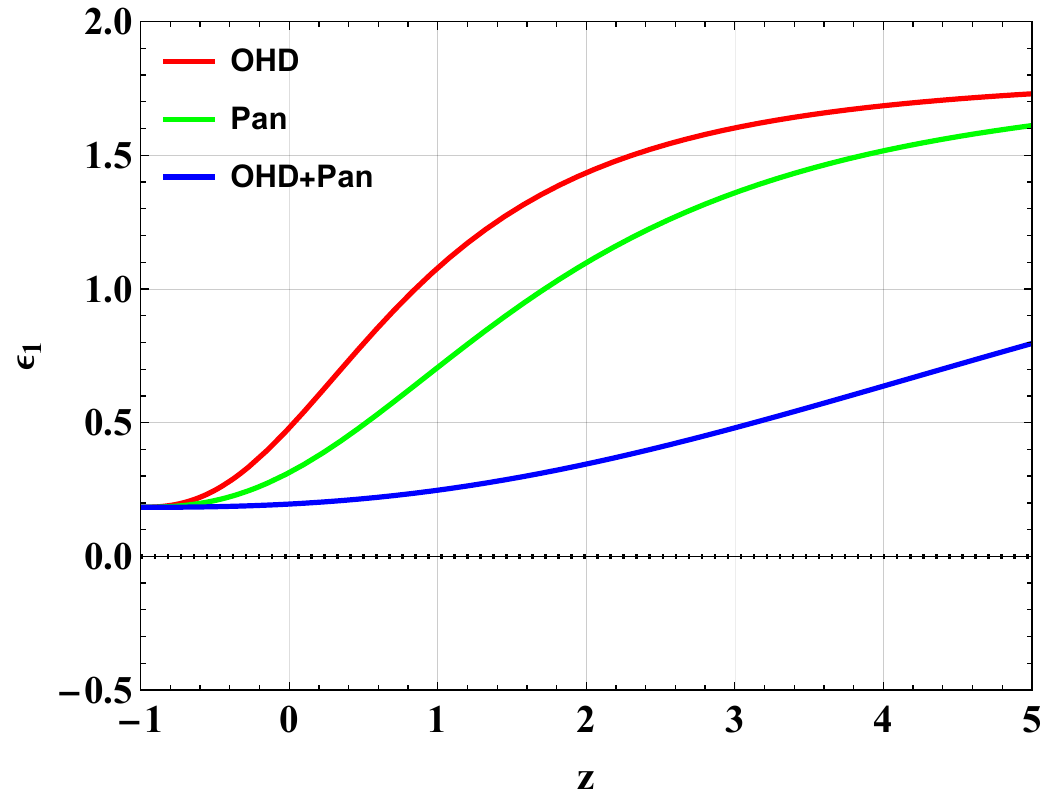}}\hfill
	\subfloat[]{\label{ssrlp}\includegraphics[scale=0.5]{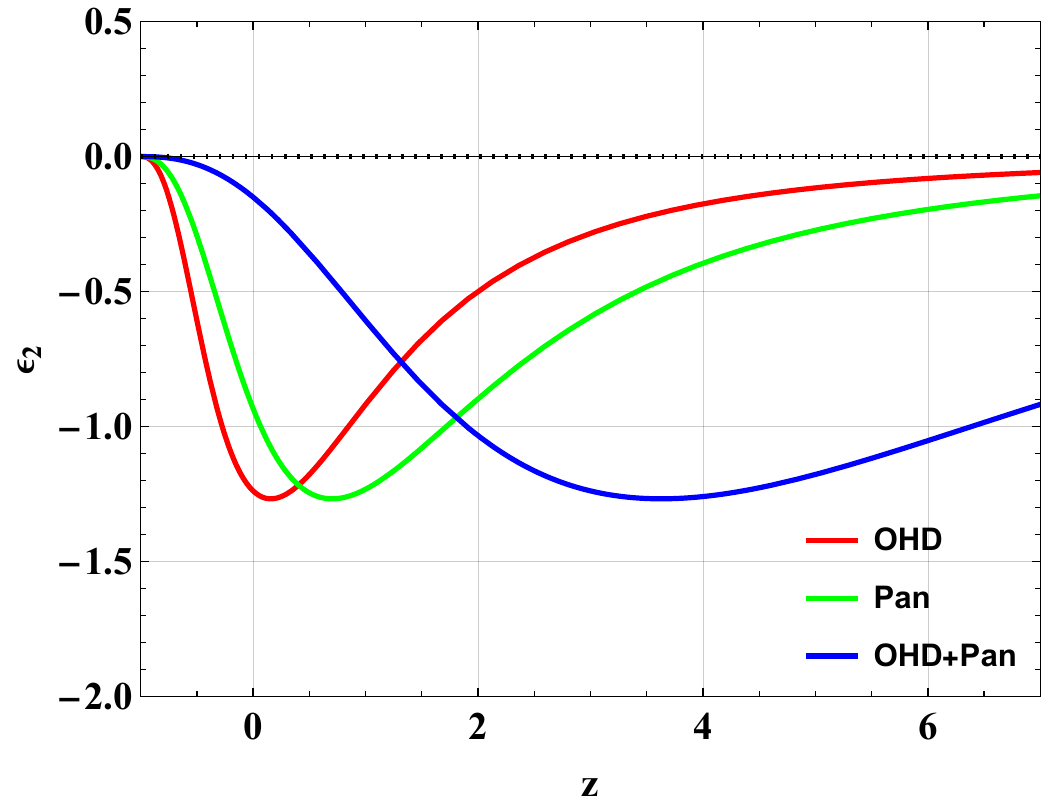}}
	\caption{ The evolution of the $ \epsilon_1 $ and $ \epsilon_2 $.}
    \label{slrlps}
\end{figure}

Using the equations,
\begin{equation}\label{40}
   \dot{H} = -(1+z) H(z) \frac{dH}{dz}
\end{equation}
and
\begin{equation}\label{41}
   \ddot{H} = \left(1+z\right)^2 \left(H(z)\right)^2 \frac{d^2H}{dz^2} + 2\left(H(z)\right)^2 (1+z) \frac{dH}{dz} + \left(1+z\right)^2 H(z) \left(\frac{dH}{dz}\right)^2 - (1+z) \left(H(z)\right)^2 \frac{dH}{dz},
\end{equation}
and $ H $ from Eq. (\ref{22}) in Eqs. (\ref{38}) and (\ref{39}), we obtain the values of both slow-roll parameters in terms of $ z $. The graphs are depicted in Fig. \ref{slrlps} with the help of the obtained values.

The first slow-roll parameter $ \epsilon_1 $ quantifies how flat the potential $ V(\phi) $ is, i.e., this parameter measures the steepness of the potential, and the second slow-roll parameter $ \epsilon_2 $ measures the curvature of the potential. The condition for inflation regarding the first slow-roll parameter $ \epsilon_1 $ can be taken as $ \epsilon_1 << 1 $, and the inflation will continue as long as the second slow-roll parameter $ \epsilon_2 $ satisfies $ |\epsilon_2| << 1 $.

Figs. \ref{fsrlp} depict that the first slow-roll parameter satisfies the condition $ \epsilon_1 << 1 $ as $  z \to -1 $, and the second slow-roll parameter obeys the condition $ |\epsilon_2| << 1 $ as $  z \to -1 $ for all the observational datasets. Thus, it can be said that the slow-roll inflation sustained for those values of $ z $ at which both the conditions $ \epsilon_1 << 1 $ and $ |\epsilon_2| << 1 $  because  $ \lim_{z \to -1}\epsilon_1,~|\epsilon_2| =0 $. These conditions imply that the inflaton field rolls slowly enough so that the universe undergoes sufficient inflation and solves the horizon and flatness problems.

\begin{figure}\centering
	\includegraphics[scale=0.51]{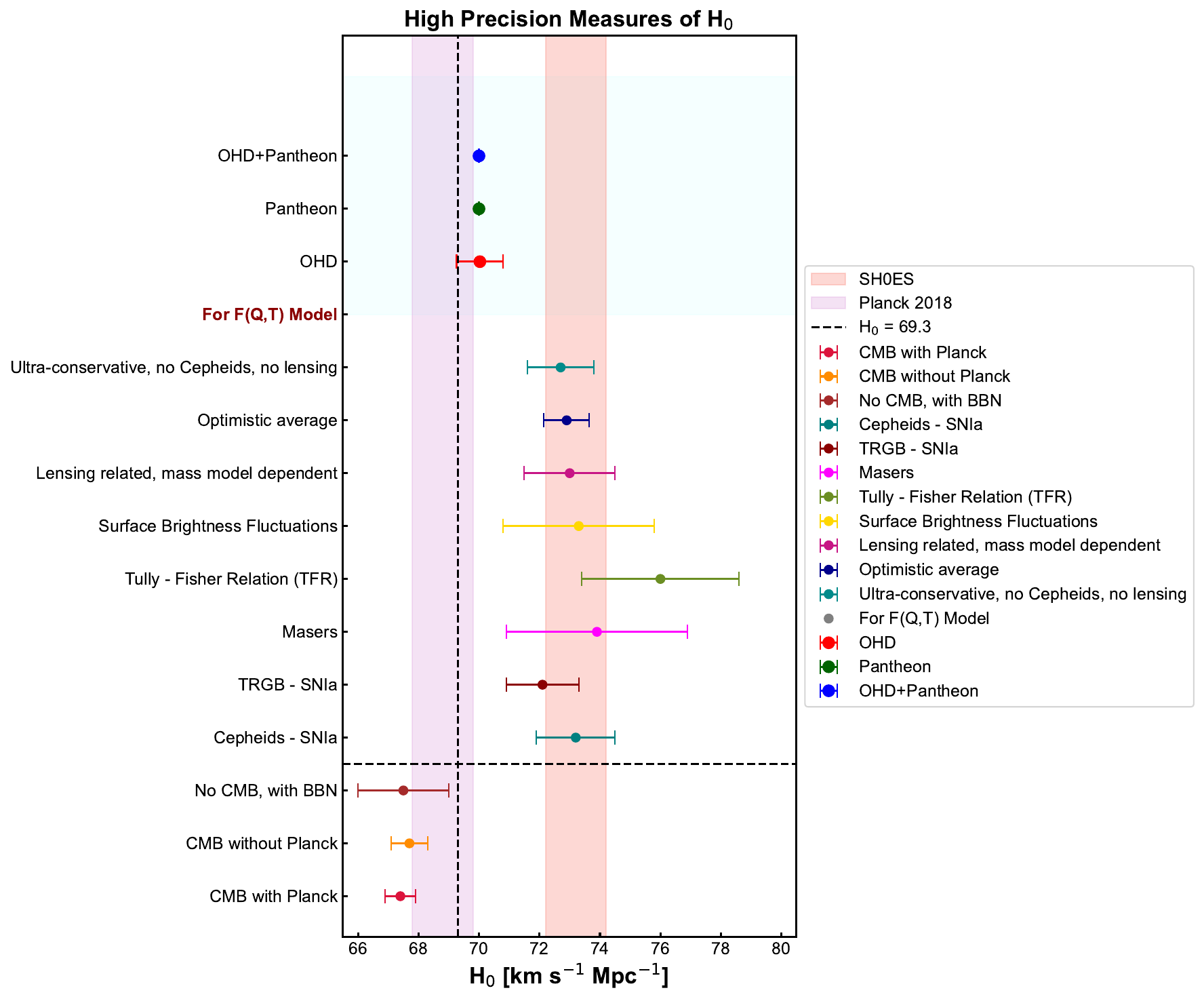}
		\caption{ The recent approximations of the $ H_0 $-tension evaluated from the OHD, $ Pantheon $, and OHD + $ Pantheon $ datasets respectively.}
		\label{whisk}
\end{figure}

\begin{table}
    \caption {\bf Summary of the best-fit value for the Hubble parameter $ H_0 $ of our model in comparison with the models given in the tables}
\begin{center}
\label{table4}
\begin{tabular}{l l l l c c} 
\hline\hline
     \\ 
  { Model}  &  ~~~~~{ Datasets}~~~~~~~~~~~~~~~~& ~~~~ $ H_0 $
        \\
        \\
        \hline
        \\
    $ f(Q,T) $   &  ~~~~~$ H(z) $+Pantheon &~~~ $ 69.99999^{+0.00011}_{-0.000087} $
    \\
       \\
         \hline
        \\
    $ f(R, T) $   &  ~~~~~$ H(z) $+Pantheon &~~~ $ 71.3^{+0.66}_{-0.25} $  \cite{Singh:2024kez}
        \\
       \\
    $ f(R, T) $   &  ~~~~~$ H(z) $+Pantheon &~~~ $ 70.3719 \pm 0.0001 $  \cite{Shaily:2024nmy}
    \\
       \\
    S\'aez Ballester   &  ~~~~~$ H(z) $+Pantheon &~~~ $ 70.81^{+0.19}_{-0.043} $  \cite{Singh:2024zvm}
    \\
       \\
    Scale-covariant theory   &  ~~~~~$ H(z) $+Pantheon &~~~ $ 70.979^{+0.021}_{-0.0043} $  \cite{Shaily:2022enj}
        \\
       \\
       \hline
        \\
    Bianchi Type-I   &  ~~~~~$ H(z) $ ($ 46 $ points)+Pantheon &~~~ $ 70.17 $  \cite{Bhardwaj:2022lrm}
        \\
       \\
       \hline
        \\
    $ f(R, T) $   &  ~~~~~$ H(z) $+Pantheon+BAO &~~~ $ 70.39 \pm 0.16 $  \cite{Singh:2024ckh}
        \\
       \\
    $ f(R, G) $   &  ~~~~~$ H(z) $+Pantheon+BAO &~~~ $ 67.9998^{+0.0093}_{-0.0082} $  \cite{Rani:2024uah}
        \\
       \\
       \hline
        \\
    $ f(R, G) $   &  ~~~~~$ H(z) $+Pantheon+Gold+Gamma Ray Burst &~~~ $ 68.115^{+0.015}_{-0.12} $  \cite{Singh:2023bjx}
        \\
       \\
       \hline
        \\
   $ f(R, L_m) $   &  ~~~~~$ H(z) $+Pantheon+Gold+Gamma Ray Burst+BAO &~~~ $ 68.0005 \pm 0.0094 $  \cite{Singh:2024gtz}
        \\
       \\
   Rastall gravity   &  ~~~~~$ H(z) $+Pantheon+Gold+Gamma Ray Burst+BAO &~~~ $70.8749_{-0.0001}^{+0.0002}$  \cite{Singh:2024urv}
           
       \\
       \\
         \hline
        \\  
   Exp $+\log F(R) $ & ~~~~~CC H(z)+SNeIa+CMB+BAO ~~~~ &~~~ $ 68.92^{+1.63}_{-1.72} $  \cite{Odintsov:2024lid}
       \\
       \\
   Exp~$ +\log F(R)~ + $~axion  & ~~~~~CC H(z)+SNeIa+CMB+BAO ~~~~ &~~~ $ 69.0^{+1.72}_{-1.71} $  \cite{ Odintsov:2024lid}
  \\
  \\
  Exp $ F(R)  $ & ~~~~~CC H(z)+SNeIa+CMB+BAO ~~~~ &~~~ $ 68.84^{+1.75}_{-1.64} $  \cite{Odintsov:2024lid}
  \\
  \\
        \hline
        \\  
  $  F(R)~+ $ EDE   & ~~~~~CC H(z)+SNeIa+CMB+BAO ~~~~ &~~~ $ 68.93^{+1.61}_{-1.57} $ \cite{Odintsov:2023cli}
  \\
    \\
 $\Lambda$CDM ~~~~     & ~~~~~CC H(z)+SNeIa+CMB+BAO  & ~~~~$ 68.98^{+1.58}_{-1.60} $ \cite{Odintsov:2023cli, Odintsov:2024lid} 
       \\
       \\
\hline\hline  
\end{tabular}   
\end{center}
\end{table}

\section{ The concluding remarks}\label{concs}

\qquad In this work, we have explored an expansion of the third identical characterization of GR known as the symmetric teleparallel formulation. This expansion is known as the $ f(Q, T) $ theory of gravity. Its theory involves a non-minimal coupling between $ Q $ and $ T $ of the EMT expressed as $ f(Q, T) = f_1(Q) + f_2(T) $, a configuration extensively studied in the existing literature. In our analysis, we specified $ f_1(Q)=\zeta Q^2 $ with $ \zeta<0 $ and $ f_2(T)=\gamma T $. We consider the jerk parameter $ j= exp(q) $ to avoid complications in the calculation and proceed to find the appropriate solutions for several physical parameters. We use the MCMC method by executing the {\it emcee~codes} in Python using various observational datasets explained in Fig. \ref{contours}. In many significant examinations, we found two-dimensional posterior classifications of the parameters $ H_0 $ and $ \alpha $ at $ 1\sigma $ and $ 2\sigma $ confidence levels (CL) that impart constraints on these. 

The error bar trajectories of $ H(z) $, and $ \mu(z) $ concerning redshifts $ z $ reveal the extent of the compatibility between the obtained model and the $ \Lambda $CDM. These plots indicate the deflections of our model from the $ \Lambda $CDM at the early evolution. Both later coincide with $ \Lambda $CDM in divergent observational datasets (see Fig. \ref{errorp}). The recent approximations of the $ H_0 $-tension obtained from the OHD, $ Pantheon $, and OHD + $ Pantheon $ datasets can be seen in Fig. \ref{whisk}.

The current value of the parameters  $ H $, $ q $, and $ \alpha $ for distinct observations are shown in Table \ref{tabparm}. We have discussed the nature of the $ q $ for good-fit values in Fig. \ref{bq}, where $ q $ transits from deceleration to acceleration in the redshift range $ 0.84 \leq z_{tr} \leq 6.39 $. Thus, the model is accelerating at present. The jerk parameter $ j $ is positive for every redshift range and tends to $ 0.5 $ in late times, as shown in Fig. \ref{cj}. The evolution of $ \rho $ is always positive and decreases monotonically as the cosmic redshift $ z $ decreases over time (see Fig. \ref{ar}). 

The isotropic pressure $ p $ transits from positive to negative in the interval of the redshift $ 0.47 < z < 4.79 $, and shows the dust-filled universe at $ z \approx 0.47,~1.14,~4.79 $ for the observations OHD, $ Pantheon $, and OHD + $ Pantheon $, respectively. This shows the negative conduct for every constraint in late times and converges to zero at $ z \to -1 $. The model demonstrates the existence of dark energy that enforces the accelerating expansion at later times (see Fig. \ref{bp}). The EoS traverses from positive to negative, which tends to $ -1 $ as shown in Fig. \ref{cw}. The model begins with the Ekpyrotic phase $ (\omega>>1) $, Perfect fluid phase $ (0 \leq \omega \leq 1) $ , and Quintessence phase $  (-\frac{1}{3} > \omega > -1) $. Hence, the model conducts like the quintessence model at later times according as Table  \ref{tabparm2}.

In Fig. \ref{ecs}, it has been seen that the model fulfills all the criteria of ECs except SEC in the later times, which suggests the presence of dark energy at later times. In this model, the DEC is disobeyed for the higher redshift, which validates the Ekpyrotic phase of the universe, and it is satisfied belated. Specifically, the clear tendency of DEC accomplishment at low redshift is similar to what was found by Lima et al. \cite{Lima:2008sq, Lima:2008ci}. The belated defiance of the SEC is directly correlated with the current accelerated expanding nature of the universe \cite{Singh:2024urv}.

The evolutions of the quintessence-like KE and PE decrease monotonically concerning the redshift $ z $, tending to zero as $ z\to -1 $, and the quintessence-like KE decreases more rapidly than the quintessence-like PE at low redshift for all observations. The quintessence-like potential energy is violated at high redshift $ z>2 $ (see Fig. \ref{scalar-z}). Fig. \ref{slrlps} demonstrate that the slow-roll inflation sustained for those values of $ z $ at which both the conditions $ \epsilon_1 << 1 $ and $ |\epsilon_2| << 1 $. These conditions imply that the inflaton field rolls slowly enough so that the universe undergoes sufficient inflation and solves the horizon and flatness problems.

Finally, we compare our model with the other standard models in Table \ref{table4} and find that our study exhibits that the quintessence model in $ f(Q, T) $-gravity imparts viable support for comprehension of the universe's evolution. It effectively assimilates the requisite observational data and provides an understanding of the energy conditions and steadiness of the universe. The review of the model endows wide-ranging interpretations of cosmology and the plausible performance of modified gravity theories elaborating on the universe's accelerated expanding nature and comprehensive dynamics. The upcoming projects will speculate further investigations into the suggestions of these consequences and the prospective for observational accreditation.

\vskip0.2in
\section*{Acknowledgement} The authors, J. K. Singh and Shaily thank the Department of Mathematics, NSUT, New Delhi-78, India, and Bennett University, Greater Noida, India, for providing the necessary facilities where a part of this work has been completed. 

\vskip0.2in

\section*{Data Availability Statement} In this manuscript, we have used observational data as available in the literature as such our work does not produce any form of new data.

\end{document}